\documentclass[onecolumn,18pt]{article}
\usepackage{array}
\usepackage{stackrel}
\usepackage{amsmath,amsthm, amssymb}
\usepackage{hyperref}
\usepackage{graphicx}
\newtheorem{theorem}{Theorem}
\begin{document}

\title{On the Estimation of Homogeneous Population Size
in a Complex Dual-record System}

\author{Kiranmoy Chatterjee\thanks{Bayesian and Interdisciplinary Research Unit, Indian Statistical
Institute, Kolkata-700108, India; E-mail: \emph{kiranmoy07@gmail.com}} \\
Diganta Mukherjee
\thanks{Sampling and Official Statistics Unit, Indian Statistical Institute, Kolkata-700108, India; E-mail: \emph{digantam@hotmail.com}}}

\date{}
 \maketitle

\begin{abstract}
Dual-record system (DRS) (equivalently two sample Capture-recapture experiment) model with time and behavioral response variation, has attracted much attention specifically in the domain of Official Statistics and Epidemiology. The relevant model suffers from parameter identifiability problem and proper Bayesian methodologies could be helpful to overcome the situation. In this article, we have formulated the population size estimation problem in DRS as a missing data analysis under both the known and unknown directional nature of underlying behavioral response effect. Two simple empirical Bayes approaches are proposed and investigated their performances for this complex model along with a fully Bayes treatment. Extensive simulation studies are carried out to compare the performances of these competitive approaches and a real data example is also illustrated. Finally, some features of these methods and recommendations to implement them in practice are explored depending upon the availability of knowledge on the nature of behavioral response effect.
%149 words

\paragraph{}\emph{Key words:} empirical Bayes: Gibbs sampling; human population; identifiability problem; missing data; model $M_{tb}$; SEM.
%\paragraph{}\emph{AMS Classification:}
\end{abstract}

\section{Introduction}
Estimation of human population size or number of vital events occurred during a given time span is a very relevant statistical concern which includes a vast area of application in the fields of Population Studies and Epidemiology. For that we generally adopt the well-known Capture-recapture type experiment. List of individuals available from different sources on the same population are framed in a contingency table where one cell is always missing and hence the data structure is often referred as a incomplete multi-way contingency table (Fienberg, 1972). Various frequentist and likelihood approaches under suitable assumptions are discussed in the literature (\textit{see} Otis et al. (1978), Seber (1986), Chao et al. (2001)) covering most of the basic and complex models developed for Capture-recapture system. In the context of human population, same system for merging information from different lists is generally known as Multiple-record system. Correction of under ascertainment in traditional epidemiological surveillance (Chao et al. (2001)), undercount estimation in census (Wolter (1986), Elliot and Little (2005)), extent of registration for vital events (ChandraSekhar and Deming (1949), Ayhan (2010)) etc. are the primal potential application of different models of Multiple-record system. Two sources of information or Dual-record system (DRS) and Triple-record system (TRS) have extensive use as more than three sources are hardly found for human population. ChandraSekar and Deming (1949) first invoked a simple capture-recapture model for homogeneous human population, known as Lincoln-Petersen Model (or, equivalently model $M_t$) which accounts for the time (\textit{t}) variation effect and independence between capture probabilities in different samples. In this context, Bayesian approach was pioneered by Castledine (1981), Smith (1991) and later, by George and Robert (1992) on a hierarchical Bayesian $M_t$ model. But this simple model doesn$'$t work well because the underlying assumption of independence is violated in most of the situations. An efficient brief review is done by Brittain and $B\ddot{o}hning$ (2009) on the various methods available by relaxing the independence assumption and associated comparative study undertaken in the DRS context. However, for a reasonably homogeneous population, $M_t$ is not at all appropriate due to list dependence. A more complicated model $M_{tb}$ nicely characterizes the list dependency by a parameter $\phi$ ($\in R^{+}$), called \emph{Behavioral Response Effect}. In demographic context, usually $\phi>1$ which implies population is recapture prone. But for a population with sensitive characteristics, such as drug users, population with Common Congenital Anomaly disease etc., $\phi<1$. In this present article, the model $M_{tb}$ is of interest in the DRS of homogeneous human population with or without the availability of prior directional knowledge on $\phi$. When such information is available, one can hope that performance should improve for any suitable method.
\paragraph{}
Otis et al. (1978) addressed the identifiability problem in $M_{tb}$ when number of sampling occasions ($T$) is greater than or equal to two. Different approaches using martingle theory (Lloyd, 1994), quasi-likelihood inference (Chao et al., 2000) are proposed to solve the problem for $T\geq3$ under the assumption considered by Otis et al. (1978). Yang and Chao (2005) developed an efficient approach using Markovian theory for $T\geq3$ without any further assumption on the basic $M_{tb}$ model. Bayesian paradigm could be helpful to reasonably overcome the non-identifiability burden with a proper subjective choice. But subjective prior elicitation are not generally encouraged on the ground of robustness. Lee and Chen (1998) applied the Gibbs sampling idea to overcome the identifiability problem in the original $M_{tb}$ model but they did not use recapture data and estimates became unstable and prior sensitive (\textit{see} Lee et al. (2003)). Later, Lee et al. (2003) considered the assumption of Otis et al. (1978) and applied noninformative priors to all model parameters except $\phi$ whose prior was chosen by a trial-and-error method and they came up with a fully Bayesian solution using MCMC. Though, their method was built for any $T\geq2$, empirical study as well as real data application were exercised in the spirit of multiple lists ($T\geq3$) problem which is common in animal abundance estimation. Hence, there is no identifiability issue and more recapture information is expected. Their estimator performs better than the conditional MLE (exists only when $T\geq3$) when prior is chosen properly and amount of recapture information is more. However, we think that the potential of the fully Bayesian method proposed by Lee et al. (2003) should be investigated for this present complex DRS situation which is not attempted earlier. As per our knowledge, the only relevant non-Bayesian estimator is due to Nour (1982) under the assumption of $\phi>1$ in an equivalent setup.
%For alternative treatments of the relevant identifiability issues in the models with heterogeneous capture probabilities in this context, see Link (2003), Holzmann et al. (2006), Farcomeni and Tardella (2012).
\paragraph{}
Therefore, the present condition motivate us to consider the problem of estimating \textit{N} from model $M_{tb}$ in DRS framework (i.e. when \emph{T}=2) with suitable Bayes and empirical Bayes strategies. In section 2, the DRS data structure along with relevant models \emph{$M_{t}$} and \emph{$M_{tb}$} are presented. We pose the said problem as a missing data problem in section 3 since one cell in contingency table (\textit{i.e.} $x_{00}$ in Table 1) is unobserved. Therefore, two empirical Bayesian approaches are discussed in addition to a fully Bayesian method. Observing some complexity in convergence and plausible inefficiency in Bayes method with non-informative priors in some situations, we reformulate EM-within-Gibbs [EWiG] type algorithm (\textit{see} Boonstra et al. (2013)) in our context and also develop another empirical Bayes method, namely Stochastic EM-within-Gibbs [SEMWiG] (details in section 3). The last one is shown to have certain advantages over the other two approaches mentioned. Non-informative or constant priors are used as far as possible for all methods. A connection has been built between DA and Lee et al. (2003) method through prior selection and inplementation. Another goal of this article is to compare the performances of DA and other two empirical Bayesian methodologies under the common roof of the incomplete two-way contingency table. Extensive empirical comparison of these approaches are illustrated in section 4.1, over different hypothetical populations. In section 4.2, a real data application on death records is presented. Finally in section 5, we summarize our findings and enumerate the best possible estimation rule depending upon the availability of directional knowledge on behavioral response effect $\phi$.

\section{Dual-record System (DRS): Preliminaries}
The idea of dual counting of certain human population size is equivalent to the very popular capture-recapture sampling approach in wildlife management. ChandraSekhar and Deming (1949) was pioneered to use this approach for human population in estimating the number of vital events and pointed out the possible correlation exists in DRS. Consider a given human population \emph{U} of true size \emph{N}. It is believed that the census or a registration system failed to capture all events. To know the true extent of this coverage or equivalently the true \emph{N}, one collects another list independently covering the same population. Then combining the two sources of information, estimate of \emph{N} could be obtained assuming different conditions on the capture probabilities of individuals leading to different models. In this paper we will concentrate on those models which have two common assumptions - (\textit{i}) population is closed between the time of the two samples taken, (\textit{ii}) individual are homogeneous with respect to capture probabilities, i.e. $p_{i1.}=p_{1.}$ and $p_{i.1}=p_{.1}$ for \emph{i}=1, 2, $\cdots$, \emph{N}, where $p_{i1.}$ and $p_{i.1}$ are the capture probability of \emph{i}th individual in first sample (List 1) and second sample (List 2) respectively. As an example, a specialized survey known as Post Enumeration Survey (PES) is conducted particularly to estimate the coverage error in Indian census. Some countries use their regular survey as the second source, e.g. US Census Bureau uses the Current Population Survey (CPS). The individuals captured in List 1 (made from census) are matched one-by-one with the list of individuals made from second sample. We classify all the captured individuals in \emph{U} according to a multinomial fashion as in the left panel of Table 1 and this particular data structure is known as Dual-record System (DRS). $x_{10}$ denotes the number of individuals present in List 1 but not in List 2. Similarly, $x_{01}$ denotes the number of individuals not present in List 1 but present in List 2. The number of individuals listed in both samples is denoted by $x_{11}$. The last cell presents the number of missed individuals by both systems and this quantity is unknown which makes the total population size \emph{N}(= $x_{..}$) unknown. Expected Proportions or probabilities for each cell are also given in the corresponding cell position at the right panel of Table 1 and these notations will be followed throughout this paper.
\begin{table}[h]
\centering
\caption{$2\times2$ table for Dual-record-System Model}
\begin{tabular}{lcccccccccc}
% creating four columns
\hline
%inserting double-line
%   &    &     &      \\
&&\multicolumn{3}{c}{Observed sample numbers}&& &&\multicolumn{3}{c}{Expected Proportions} \\
\cline{3-5}\cline{9-11}
&&\multicolumn{3}{c}{List 2} &&&& \multicolumn{3}{c}{List 2} \\
List 1& & In & out & Total & && & In & out & Total \\
\cline{3-5}\cline{9-11}
In & &$x_{11}$ & $x_{10}$ & $x_{1.}$ & && & $p_{11}$ & $p_{10}$ & $p_{1.}$\\
Out& &$x_{01}$ & $x_{00}$ & $x_{0.}$ & && & $p_{01}$ & $p_{00}$ & $p_{0.}$\\ % [1ex] adds vertical space
\cline{3-5}\cline{9-11}
Total& & $x_{.1}$ & $x_{.0}$ & $N$ & && & $p_{.1}$ & $p_{.0}$ & $1$\\
\hline
% inserts single-lin
\end{tabular}
\label{tab:dse}
\end{table}

\subsection{Time Variation Model: $M_{t}$}
This model is simple and widely used under the capture-recapture framework. Two additional assumptions are required for this model. One is that the two lists are causally independent. The event of being included in List 1 is independent of the event of being included in List 2. Hence, the cross-product ratio satisfies $\psi=p_{11}p_{00}/(p_{01}p_{10})=1$. Another assumption is time variation in the capture probabilities, i.e., $p_{1.}\neq p_{.1}$. Then for given the data (\textbf{\underline{x}=($x_{11}$, $x_{10}$, $x_{01}$)}), the associated likelihood for model parameter $\theta=(N,p_{1.},p_{.1})$ is
\begin{equation}
L_{t}(\theta|\textbf{\underline{x}}) \propto \frac{N!}{x_{11}!x_{01}!x_{10}!(N-x_0)!}p_{1.}^{x_{1.}}p_{.1}^{x_{.1}}(1-p_{1.})^{N-x_{1.}}(1-p_{.1})^{N-x_{.1}}.
%\label{eq:luminosity}
\end{equation}
The corresponding maximum likelihood estimates are \\
\[ \hat{N}_{M_t} = \frac{x_{.1}.x_{1.}}{x_{11}},\hspace{0.2in} \hat{p}_{1.,M_t}=\frac{x_{11}}{x_{.1}} \hspace{0.2in} \mbox{and} \hspace{0.2in} \hat{p}_{.1,M_t}=\frac{x_{11}}{x_{1.}}. \]
Typical capture-recapture theory under this model says that fraction recaptured among the
second sample estimates the fraction of the whole population caught
the first time. ChandraSekar et al. (1949) pointed out the possibility of existence of lists dependence which make the corresponding estimate as biassed one. Two types of dependence may be identified. The first type is source correlation which results from either planned or unplanned communication between the two lists. The second type is either due to respondent bias or due to the nature of the events themselves and it may arise even if the two lists are operationally independent. However, we call this second type dependence as causal dependence and will discuss two basic models under this dependence.

\subsection{Time-Behavioral Response Variation Model: $M_{tb}$}
Causal independence assumption is criticised in surveys and censuses of human populations. The concern is that an individual's capture probability in second list may change in response to being captured in the first list. An individual who is captured in first attempt may have more (or less) chance to be included him/herself in the second list than the individual who has not been captured in first attempt. This change is grossly known as behavioral response variation. When this behavioral change in response is relevant in addition to the model $M_{t}$, one will get the relatively complex model $M_{tb}$. To model this situation for a homogeneous population, we have from Table 1 that Prob(An individual is captured in List 1) = $p_{1.}$ and let us denote Prob(\emph{i}th individual is captured in List 2 $|$ he/she is not captured in List 1)= $p_{01}/p_{0.}$ = $p$ (say). But this model suffers from identifiability problem due to the number of sufficient statistics being less than the number of relevant parameters as addressed by Otis et al. (1978). Following a frequently used assumption in capture-recapture literature, here we also assume that the recapture probability at second sample, \emph{c}, is proportional to the $p$ and hence, $c=\phi p$. The constant $\phi$ is referred as \textit{behavioral response effect}. Though this assumption can not avoid the identifiability problem as number of parameters is not reduced, still we consider it for our convenience. Hence our likelihood for model parameter $\theta=(N,p_{1.},p,\phi)$ becomes
\begin{equation}
L_{tb}(\theta|\textbf{\underline{x}}) \propto \frac{N!}{(N-x_0)!}\phi^{x_{11}}p_{1.}^{x_{1.}}p^{x_{.1}}(1-p_{1.})^{N-x_{1.}}(1-p)^{N-x_{0}}(1-\phi p)^{x_{10}}.
\end{equation}
Therefore individuals in the population \textit{U} are said to be \emph{recapture prone} only if $\phi>1$ and they are called \emph{recapture averse} only if $\phi<1$. When $\phi=1$, then (2) will be equivalent to (1) i.e. $M_{tb}\Rightarrow M_{t}$ as conditional $p$ will be equal to unconditional $p_{.1}$. Again, when $\phi \neq 1$ but $p_{1.}=p_{.1}$, then $M_{tb}\Rightarrow M_{b}$, another model not relevant for human population. In the next two sections, we demonstrate three different versions of Gibbs samplers in order to obtain Bayes and Empirical Bayes estimates of \emph{N} under the model $M_{tb}$.

\section{General Bayesian Framework}
From section 2, it is clear that model $M_{tb}$ has a parameter identifiability problem and Bayesian paradigm could be helpful to get rid of it. Here we first reformulate the \textit{N} estimation problem in a slight different way. Population size (\textit{N}) estimation in Multiple-record system problems can be viewed as estimation of missing data (see Fienberg and Vallier (2009)). Several fully and partially Bayesian methods, which can be used in this missing data context, will be discussed here in a general setup. Henceforth we illustrate the implementation of each method in the context of population size estimation under the model $M_{tb}$.
\paragraph{}
Let us consider a typical missing data situation where the complete data can be classified into observed ($U^{obs}$) and missing ($U^{mis}$) components. We may parameterize the whole model with k-component interest parameter $\theta=(\theta_1,\theta_2,\ldots,\theta_k)$ and nuisance parameter $\psi$ (may also be a vector). A subset $\theta^S$ of $\theta$ denotes missing quantity $U^{mis}$ and rest part $\theta_{-S}=\theta\setminus\theta^S$ is model parameter of interest. Now we discuss two different variants of the Gibbs sampler relevant to our model - (a) Data Augmentation [DA] and (b) EM-within-Gibbs [EWiG]. Here 'variant' refers to strategies for fitting the model given the posterior distribution. DA (Tanner and Wong, 1987) is a standard Bayesian approach to missing data. The second one, EWiG (Boonstra et al., 2013) is an Empirical Bayes approach which is an expanded version of Monte Carlo Expectation–Maximization algorithm [MCEM] (Wei and Tanner, 1990). DA is a very standard technique allowing priors to all unobserved random quantity where as EWiG determines the posterior distribution of all the unobserved quantities of the model except the nuisance parameter. Moreover, we sketch another alternative approach, which we call \emph{Stochastic EM-within-Gibbs} [SEMWiG], that possess certain advantages over two other methods.
\paragraph{}
In the following course of discussions, we present three approaches (DA, EWiG and SEMWiG) in a general missing data model with $\theta$ and $\psi$ parametrization as formulated above. Each approach has its own posterior of interest depending upon whether non-empty $\psi$ is considered to be random or not. Particularly, for DRS, $U^{obs}=(x_{11},x_{12},x_{21})$ and we may replace $U^{obs}$ with \textbf{\underline{x}} shortly. Moreover, size of unobserved individual in both the samples ($x_{00}$) makes \textit{N} unknown. Hence, one can think $x_{00}=N-x_0$ as missing data. Unlikely to the general missing data analysis, missing cell value ($x_{00}$) is certainly a parameter of interest in any problem involving DRS. Therefore, we consider $\theta^S=x_{00}$ or equivalently, $\theta^S=N$ and $(\theta_{-S}, \psi)=(p_{1.}, \phi, p)$. In the current article. our intension is to always use non-informative or uniform priors for all unobserved random quantities in the model so that subjectiveness would be as minimum as possible. Hence in that context, the posterior of interest becomes $[\theta|\textbf{\underline{x}},\psi]$ or $[\theta|\textbf{\underline{x}},\psi,\gamma]$  when $\psi$ is assumed to be nonrandom. One may like to model $x_{00}$ by a prior density with a unspecified hyperparameter $\gamma$. For example, poisson prior Poi($\gamma$) is a good choice. The required marginal posterior density becomes $[\theta^{S}|\textbf{\underline{x}},\psi]$ for $M_{tb}$ to estimate population size only. Since, data generation in DRS or capture-recapture experiment falls under the finite population sampling. Hence, convergence of posterior density can be studied as the population size, $N \rightarrow \infty$.

\subsection{Data Augmentation [DA]}
Data augmentation refers to strategies for constructing iterative optimization or sampling algorithms for all the unknown quantities in model. The first variant we discuss here is a fully Bayesian treatment renowned for the analysis of missing data considering all the unknown quantities in the model to be random. Hence, prior densities have to be assigned for both of $\theta$ and $\psi$. Posterior samples of them can be drawn iteratively through a suitable Gibbs sampling strategy which is certainly feasible for any model of interest. Tanner and Wong (1987) developed this stochastic version of data augmentation (DA) to make simulation simple and straightforward. If $\pi(\psi|\theta,U^{obs})$ and $\pi(\theta_i|\theta_{-i},\psi,U^{obs})$, for $i=1,2,\ldots,k$ are the resultant conditional posterior distributions, then stochastic DA is performed in the following way:
\\
\\
\texttt{Step 1: Set $t=0$ and initialize $\theta^{(0)}$.\\
Step 2: Generate $\psi$ and \{$\theta_i;i=1,2,\ldots,k$\} from $\pi(\psi|\theta^{(t)},U^{obs})$ and $\pi(\theta_i|\hat{\theta}_{-i}^{(t)},\psi,U^{obs})$ \\respectively.\\
Step 3: Update $\theta^{(t)}$ with $\theta^{(t+1)}=\{\theta_i;i=1,2,\ldots,k\}$.\\
Step 4: Repeat the last two steps until the convergence of $\{\theta^{(t)}\}_{t\geq 0}$.} \\
\\
Finally, samples after burn-in period are considered to be generated from the targeted marginal posterior $\pi(\theta|U^{obs})$. Lee et al. (2003) performed this strategy by assigning priors to all model parameters. Hence, the Bayesian method used by Lee et al. (2003), for estimating population size from multiple capture-recapture system, is a stochastic data augmentation procedure.
\paragraph{}
\textbf{\textit{Implementation.}} We take $\theta^{S}=N-x_{0}=x_{00}$, $\theta_{-S}=(p_{1.}, \phi, p)$ i.e. $\psi$ remains empty. In full Bayesian analysis, priors on \textit{N} and ($p_{1.}, \phi, p$) may be assigned independently as $\pi(N)\pi(\theta)=\pi(N)\pi(p_{1.})\pi(\phi)\pi(p)$ and hence separate conditional posterior densities for \textit{N}, $p_{1.}$, $\phi$ and $p$  are obtained as follows
\begin{eqnarray}
\pi(x_{00}|,p_{1.},p,\textbf{\underline{x}}) & \propto & \frac{N!}{(N-x_0)!}((1-p)(1-p_{1.}))^N\pi(N),\\
\pi(p_{1.}|N,\textbf{\underline{x}}) & \propto &  p_{1.}^{x_{1.}}(1-p_{1.})^{N-x_{1.}}\pi(p_{1.}),\\
\pi(\phi|N,p,\textbf{\underline{x}}) & \propto &  \phi^{x_{11}}(1-\phi p)^{x_{10}}\pi(\phi),\\
\pi(p|N,\phi,\textbf{\underline{x}}) & \propto &  p^{x_{.1}}(1-p)^{N-x_0}(1-\phi p)^{x_{10}}\pi(p),
\end{eqnarray}
It is clear that known expressions of conditional posterior densities can be available only for $N$, $p_{1.}$ and $\phi$ and to draw samples from (6), any efficient sampling strategy may be adopted such as adaptive rejection sampling. Different priors, both non-informative and informative, can be considered for all the random quantities. However, as we are here also interested to use non-informative or constant prior like Lee et al. (2003), here we follow their prior assignment. If we follow their approach for dual system, adaptive rejection sampling is needed only to generate \textit{p} from (6) for any prior on \textit{p}. Lee et al. took non-informative priors $\pi(p_{1.})=Unif(0,1)$, $\pi(p)=Unif(0,1)$ and Jeffrey's prior for $x_{00}$ or equivalently $\pi(N)\propto N^{-1}$. A flat prior $\pi(\phi)$=Unif($\alpha$, $\beta$) is chosen though this is no longer non-informative as we need to specify its range [$\alpha$, $\beta$] but rather a flat prior. Thus full conditional posteriors for all the unobserved random quantities are
\begin{eqnarray}
\pi(x_{00}|,p_{1.},p,\textbf{\underline{x}}) & \propto &  \mbox{Neg. Binomial($x_0$,$\mu$)},\\
\pi(p_{1.}|N,\textbf{\underline{x}}) & \propto &  Beta(x_{1.}+1,N-x_{1.}+1),\\
\pi(p|N,\phi,\textbf{\underline{x}}) & \propto &  p^{x_{.1}}(1-p)^{N-x_0}(1-\phi p)^{x_{10}},\\
\pi(\phi|N,p,\textbf{\underline{x}}) & \propto &  \mbox{Gen. Beta-I}(x_{11}+1,x_{10}+1,1,p)\times \mathcal{I}_{[\alpha,\beta]}(\phi),
\end{eqnarray}
where $\mu=1-((1-p)(1-p_{1.}))$ and $\mathcal{I}_{[\alpha, \beta]}(\phi)$ is an indicator function for $\phi \in \Phi$. Draws of unobserved $x_{00}$ (or, $N-x_0$) and $\theta$ are sequentially made from their conditional posteriors and iterations will ultimately yield draws from the true marginal posterior distributions after certain burn-in period. To fix the burn-in period, we adopt the $\hat{R}^{1/2}$ technique suggested by Gelman (1996) following Lee et al. (2003). If no other information on $\phi$ is available then choosing suitable prior is not at all easy. Lee at al. (2003) adopted a trial-and-error procedure. They opt for such $\alpha$ and $\beta$ values for which the range of the posterior credible interval for $\phi$ is not too close to either side of the prior limits. This procedure works well when there is a large amount of recapture information and ($\alpha$, $\beta$) can be chosen from the experiences of experts. Usually for human population, high capture probabilities may be attained but number of samples is usually small (not more than three). In Lee et al. (2003), no investigation has been carried out separately in their article for given \textit{trap-shy} (i.e. $\phi<1$) and \textit{trap-happy} (i.e. $\phi>1$) populations. Since we are also interested in the situations where directional knowledge on behavioral effect is available, so will start with the first trial with $\alpha=0$ and $\beta=1$ naturally for recapture-averse population and $\alpha=1$ and $\beta=2$ or 3 for recapture-prone population and observe the performance of DA approach under this available knowledge.

\subsection{EM-within-Gibbs [EMWiG]}
Instead of using full Bayesian method, if some model parameters are restricted to be updated by any consistent likelihood or frequentist method, then it will certainly reduce the variance of the generated sequence. Boonstra et al. (2013) presents a generic approach EM-within-Gibbs (EWiG) by expanding MCEM approach (developed by Wei and Tanner (1990)) in the presence of an unknown nuisance hyperparameter. Considering the idea of from EWiG method, we adopt an equivalent empirical Bayes version of Gibbs sampler in our present context where prior densities are assigned on the interest parameter $\theta$ only and nuisance parameter $\psi$ (may also be a vector) remains to be estimated by a classical approach, such as mle. Let us denote $ln(U^{obs}|.)$ as log-likelihood function for given data $U^{obs}$. For a given point estimate $\hat{\psi}$, usually an mle under the marginal $ln(U^{obs}|\psi)$, Gibbs sampler generates \textit{M} draws from $\pi(\theta|\hat{\psi},U^{obs})$. To obtain the mle, computation of marginal log-likelihood $ln(U^{obs}|\psi)$ is required, which may include high-dimensional integration. However, one can write the marginal likelihood for $\psi$ as $[U^{obs}|\psi]=[\theta,U^{obs}|\psi]/[\theta|\psi,U^{obs}]$. EM algorithm can automatically produce the marginal mle of $\psi$ after taking expectation on $ln[U^{obs}|\psi]$ or equivalently, on $ln[\theta,U^{obs}|\psi]$ with respect to $[\theta|\psi,U^{obs}]$. E-step nicely engaged the conditional posterior density of $\theta$, $\pi(\theta|\psi,U^{obs})$ and then at M-step, $E[\theta,U^{obs}|\psi]$ or its empirical version is maximized with respect to $\psi$. A monte carlo treatment for calculating the empirical EM is
\begin{equation}
\hat{\psi} =\hspace{0.1in} \stackrel[\psi\in\Psi]{}{argmax}  M^{-1}\sum_{j=1}^{M}l(\theta^{(j)},U^{obs}|\psi),
\end{equation}
which is similar to the Monte Carlo EM (Wei and Tanner, 1990) approach. Considering $\theta^{S}$ as missing random quantity and $\theta_{-S}$ as parameter of interest, we sketch an Gibbs sampling algorithm with one Expectation-Maximization step for our current problem of interest in a generic missing data context following, but not identical to, EWiG (Boonstra et al., 2013). The whole Gibbs sampler proceeds as follows to obtain the approximate posterior densities.
\\
\\
\texttt{Step 1: Set $t=0$ and initialize $\psi^{(0)}$.\\
 Step 2: Generate \{$\theta^{(j)}=(\theta^{S(j)},\theta_{-S}^{(j)}); j=1(1)M$\} by iteratively simulating from $\pi(\theta^S|\theta_{-S},\psi^{(t)},U^{obs})$ and $\pi(\theta_{-S}|\theta^S,\psi^{(t)},U^{obs})$.\\
Step 3: \textit{Expectation and Maximization}. Obtain $\psi^{(t+1)}$ by updating $\psi^{(t)}$ using (3.9).\\
Step 4: Repeat the above two steps until the convergence of $\{\psi^{(t)}\}_{t\geq 0}$.}\\
\\
Since, the nuisance parameter $\psi$ is estimated by EM, hence the sequence $\{\psi^{t}\}$ converges to the mle of $\psi$. At the convergent $\{\psi^{(t)}\}$ at $\hat{\psi}$, the final sample \{$\theta^{(j)}=(\theta^{S(j)},\theta_{-S}^{(j)}); j=1(1)M$\} is drawn following Step 2. From this final sample we would obtain the estimate of posterior density $\pi(\theta_{-S}|\psi,U^{obs})$ and $\pi(\theta^{S}|\psi,U^{obs})$ respectively as
\begin{equation}
\hat{\pi}(\theta_{-S}|\hat{\psi},U^{obs})= M^{-1}\sum_{j=1}^{M}\pi(\theta_{S-}|\theta^{S(j)},\hat{\psi},U^{obs})
\end{equation}
\begin{equation}
\mbox{and $\hat{\pi}(\theta^{S}|\hat{\psi},U^{obs})$= $M^{-1}\sum_{j=1}^{M}\pi(\theta^S|\theta_{-S}^{(j)},\hat{\psi},U^{obs})$}.
\end{equation}
In any such empirical Bayes procedure, a fundamental concern is that how one can consider $\hat{\pi}(\theta_{-S}|\hat{\psi},U^{obs})$ and $\hat{\pi}(\theta^{S}|\hat{\psi},U^{obs})$ as estimates of $\pi(\theta_{-S}|\psi,U^{obs})$ and $\pi(\theta^{S}|\psi,U^{obs})$ respectively. The following theorem gives us the conditions under which the acceptability of $\hat{\pi}(\theta'|\hat{\psi},U^{obs})$ as an estimate of $\pi(\theta'|\psi,U^{obs})$ for $\theta'=\{\theta_{-S},\theta^{S}\}$ are established.
 \begin{theorem}
 \textit{If $\hat{\psi}\stackrel{a.e.}{\rightarrow} \psi$ as $N\rightarrow\infty$ with respect to the marginal density $m(U^{obs}|\psi)$ and $\pi(\theta'|\psi,U^{obs})$ is continuous in $\psi$ for $\theta'=\{\theta_{-S},\theta^{S}\}$, then
\begin{enumerate}
  \item $\int_{\Theta_{-S}}\mid M^{-1}\sum_{j=1}^{M}\pi(\theta_{-S}|\theta^{S(j)},\hat{\psi},U^{obs})-\pi(\theta_{-S}|\psi,U^{obs})\mid d\theta_{-S}\hspace{0.1in} \stackrel{a.e.}{\rightarrow} 0$,
  \item $\int_{\Theta^{S}}\mid M^{-1}\sum_{j=1}^{M}\pi(\theta^{S}|\theta_{-S}^{(j)},\hat{\psi},U^{obs})-\pi(\theta^{S}|\psi,U^{obs})\mid d\theta^{S}\hspace{0.1in} \stackrel{a.e.}{\rightarrow} 0$,
\end{enumerate}
as $M, N\rightarrow\infty$.}
 \end{theorem}
\paragraph{}
The above formulated algorithm in in a general framework so that $\theta$ or any subset of $\theta$ may be thought as missing ($\theta^{S}$). If $\theta$ refers to missing unobserved quantity only and $\psi$ denotes the rest of the parameters in the model then above model becomes the originally proposed MCEM. In this way, the proposed algorithm can be characterized as EWiG-type algorithm.
\paragraph{}
\textbf{\textit{Implementation.}} When population size estimation from incomplete $2\times2$ contingency table (\textit{see} Table 1) is viewed as a missing data problem, then our model of interest is not exactly same as Boonstra et al. (2013). To apply the above EWiG-type algorithm to analyse the DRS data supposed to be generated from $M_{tb}$, we take $\theta^{S}=N-x_{0}$, $\theta_{-S}=(p_{1.},\phi)$ and $\psi=p$. With this consideration of $\theta_{-S}$ and $\psi$, we call this method as EWiG-I. Priors on $N$ and $p_{1.}$ are assigned as stated earlier in DA method (section 3.1). Same flat prior Unif($\alpha$, $\beta$) is chosen for $\phi$ but here we give a plan to assign these prior limits strategically. Since $c=\phi p<\phi$, one can always think of \emph{c} as a natural lower limit of $\alpha$. Again $\phi p=c<1$ implies $\phi<p^{-1}$. So, when it is correctly known that the population is recapture prone, then we recommend to set $\alpha$ to 1 and $\beta=p^{-1}$. When the population is recapture averse, $\alpha=c$ and $\beta=1$ will be the natural choice. Hence according to Step 2 in the EWiG-I algorithm, iterative simulation is performed between the conditional posterior densities (7), (8) and (10). Our approach is not subjective as we consider the natural bounds for the limits ($\alpha$, $\beta$) of flat uniform priors depending upon the availability of correct knowledge about the direction of $\phi$ value. In practice, $c$ is replaced by its mle $\hat{c}=x_{11}/x_{1.}$. Using updated sample \{$\theta^{(j)}=(N^{(j)},p_{1.}^{(j)},\phi^{(j)}); j=1(1)M$\}, $p$ is estimated by maximizing the empirical average of log-likelihoods $l(p|\theta^{(j)},\textbf{\underline{x}})$ for $j=1(1)M$ following (11).
\paragraph{}
When no directional information on $\phi$ is available, we suggest to use conjugate prior density $\pi(\phi|p)$ = Generalised Beta Type-I ($u$, $v$, $rate=p$) which produces conditional posterior density of $\phi$ as
\begin{equation}
\pi(\phi|p,\textbf{\underline{x}}) \propto  \mbox{GB-I ($x_{11}+u$, $x_{10}+v$, $1$, $rate=p$)}\\
\end{equation}
and hence corresponding log-likelihood for $p$ becomes $ln[p^{x_{.1}}(1-p)^{N-x_0}(1-\phi p)^{x_{10}+v-1}]$. If we take non-informative prior with (\textit{u}, \textit{v})=(0, 0), that leads to eliminate the influential terms in (14) and conditional log-likelihood for $p$. If we take (\textit{u}, \textit{v})=(1, 1), prior $\pi(\phi|p,\textbf{\underline{x}})$ simply reduces to flat uniform distribution Unif($0,p^{-1}$). Another strategy can be adopted which allow to consider $\theta_{-S}=p_{1.}$ only and $\psi$ also includes $\phi$ as nuisance parameter along with $p$. We call this EWiG strategy as EWiG-II and it can be noted that it is very close to the original MCEM approach as samples from posterior density (8) will eventually converge to its mle due to flat prior $\pi(p_{1.}|p)=Unif(0,1)$ is assigned for $p_{1.}$ on its whole domain $(0,1)$.
\paragraph{}
Since in this application, $\theta^{S}$ is equivalent to \textit{N} which is also the index in the asymptotic result in any finite population inference. Hence, the theorem 1(2) does not hold here exactly. However, we have a result about the tail convergence of $\hat{\pi}(N|\hat{p},\textbf{\underline{x}})$ to $\pi(N|p_0,\textbf{\underline{x}})$ for the true value $p_0$ of \textit{p}.
\begin{theorem}
If $\hat{p}\stackrel{a.e.}{\rightarrow} p_0$ as $N\rightarrow\infty$ with respect to the marginal density $m(\textbf{\underline{x}}|p)$ and $\pi(N|p,\textbf{\underline{x}})$ is continuous in $p$, then $\hat{\pi}(N|\hat{p},\textbf{\underline{x}})$ is right tail equivalent to the marginal posterior density $\pi(N|p_0,\textbf{\underline{x}})$ for sufficiently large $M$.
\end{theorem}
Proof of the above theorem is in \textit{Appendix B}. Theorem 2 also holds when $\phi$ is considered as nuisance parameter along with \textit{p} in EWiG-II.

\subsection{Stochastic EM-within-Gibbs [SEMWiG]}
Constructing data augmentation schemes that result in both simple and fast as well as efficient algorithms is a matter of art and also depends greatly on the models being considered. We now propose and discuss a very interesting alternative Gibbs sampling technique which is simpler than EWiG and hence takes lesser time in computation. This method is a stochastic extension of EM within Gibbs and we call it as Stochastic EM-within-Gibbs [SEMWiG]. If we put \textit{M}=1 in original MCEM, then we would have Stochastic Expectation-Maximization (SEM) procedure. Though connection between SEM and MCEM seems very simple but their underlying philosophy is different. In SEM, a stochastic imputation is done for the unobserved missing quantity $\theta^{S}$ and therefore, $\theta_{-S}$ is estimated from complete data log-likelihood. However MCEM, as we have seen earlier, replaces intractable computation of the conditional expectation of the log-likelihood of the complete data using a monte carlo approximation. Celeux (1985) advocated that both the MCEM and SEM are the stochastic perturbations of the discrete-time dynamic system generated by Expectation-Maximization (EM) procedure. In addition to that Celeux (1985) also provided several evidences on the preference of SEM over EM. Moreover, EM algorithm is sensitive to choice of starting values. SEM algorithm may accelerate the converegence (Celeux, 1985) and it is known to be more robust to poorly specified starting values (Gilks et al., 1996). SEM provides estimate of posterior density for $\theta^{S}$ but point estimates for other model parameters. However in our study, SEMWiG is designed with such a flexibility that one can have posterior density estimate also for $\theta_{-S}$ by suitably modifying the iteration steps. The whole Gibbs sampler including SEM step proceeds as follows to obtain approximate posterior distributions.
\\
\\
\texttt{Step 1: Set $t=0$ and initialize $\psi^{(0)}$ and $\theta_{-S}^{(0)}$.\\
Step 2: \textit{Stochastic Imputation}. Generate pseudo-complete data by \\simulating $\theta^S$ from $\pi(\theta^S|\theta_{-S}^{(t)},\psi^{(t)},U^{obs})$ and then simulate $\theta_{-S}$ from $\pi(\theta_{-S}|\theta^S,\psi^{(t)},U^{obs})$.\\
Step 3: \textit{Maximization}. Obtain the update of $\psi^{(t)}$ as
\[ \psi^{(t+1)} =\hspace{0.1in} \stackrel[\psi\in\Psi]{}{argmax}  l(\psi|U^{obs},\theta^{S},\theta_{-S}) \]
Step 4: Repeat the above three steps until the convergence of $\{\theta^{S(t)},\theta_{-S}^{(t)}\}_{t\geq 0}$.}\\
\\
Finally, $\hat{\psi}$ is calculated by averaging over a sufficient number of $\psi^{(t)}$'s after $\psi^{(t)}$ has reached its stationary regime. Equivalently, from the Gibbs sampler $\{\theta^{S(t)}; t=h+1,\ldots \}$ after burn-in period \textit{h}, we would obtain the estimate $\hat{\pi}(\theta^{S}|\hat{\psi},U^{obs})$ of marginal posterior density $\pi(\theta^{S}|\psi,U^{obs})$. Therefore equivalent result as in Theorem 1 can be proved for SEMWiG approach under the assumption of ergodicity of SEM generated homogeneous Markov chain (which generally holds) and consistency of $\hat{\psi}$.
%The same fundamental concern is also considered in this empirical Bayes method which allow $\hat{\pi}(\theta^{S}|\hat{\psi},U^{obs})$ to be an estimate of $\pi(\theta^{S}|\psi,U^{obs})$. The following theorem gives us the conditions under which the acceptability of $\hat{\pi}(\theta^{S}|\hat{\psi},U^{obs})$ is established.
%\begin{theorem}
%If $\hat{\psi}\stackrel{a.e.}{\rightarrow} \psi$ as $N\rightarrow\infty$ with respect to the marginal density $m(U^{obs}|\psi)$ and $\pi(\theta^{S}|\psi,U^{obs})$ is continuous in $\psi$, then for any measurable set $D\subseteq\Theta^{S}$,  \\
%\[ \int_{D}\mid \hat{\pi}(\theta^{S}|\hat{\psi},U^{obs})-\pi(\theta^{S}|\psi,U^{obs})\mid d\theta^{S}\hspace{0.1in} \stackrel{a.e.}{\rightarrow} 0, \]
%as $N\rightarrow\infty$.
%\end{theorem}
\paragraph{}
Application of SEM algorithm does not result in a single value for a parameter estimate. Instead there is built-in variation, induced by the simulated data, around the estimate, and the result will be similar to that of a stationary Markov Chain Monte Carlo estimator (Gilks et al. 1996). However, our SEM motivated approach, SEMWiG, is rather different than original SEM, because in SEMWiG, we can at most think all the parameters except $\psi$ as missing for which posterior density can be estimated. Moreover, SEMWiG will take lesser time than the previously discussed EWiG as \textit{M}=1.
\paragraph{}
\textbf{\textit{Implementation.}} To implement the above stated partially Bayes techniques here also we consider $\theta^{S}=x_{00}=N-x_{0}$, $\theta_{-S}=(\phi,  p_{1.})$ and $\psi=p$. Same priors are consider for \textit{N}, $p_{1.}$ and $\phi$ as previously discussed depending upon the availability or non-availability of the directional knowledge on $\phi$. If hyperparameter is taken as in the case of poisson prior for \textit{N}, then one can adopt the said Empirical Bayes approaches by setting $\gamma=\lambda$. The initial value of parameter vector $\theta$ may come from a wide range of choices, so it might become unstable at the beginning of the process.
We produce a Gibbs sequence \{($p_{1.}^{(t)}$, $\phi^{(t)}$, $N^{(t)}$); \emph{t} = 1, 2, 3, ...\} and hence choose the burn-in period following the same $\hat{R}^{1/2}$ technique in case of DA.
%Jeffrey's prior for \emph{N} is equivalent to the Jeffrey's prior for hyperparameter $\lambda$ (i.e. $\pi(\lambda)\propto 1/\lambda $) when $\pi(N-x_0)$=Poisson($\lambda$).
\paragraph{}
\textit{Remark.} In EWiG, if $M=1$ is considered and that single imputation is the expected value of $\theta^{S}$ having conditional posterior density $\pi(\theta^{S}|\theta_{-S},\psi,U^{obs})$, then that algorithm is often called EM-type algorithm. When $ln[\psi|\theta^{S},\theta_{-S},U^{obs}]$ is linear in $\theta^{S}$ and $\theta_{-S}$, then EM and EM-type algorithm both produce the maximizer of the observed posterior (Wei and Tanner, 1990). The proposed SEMWiG is different than EM-type based approach as in SEMWiG, we are not restricted to select expected value of $\theta^{S}$. Moreover, in application to $M_{tb}$-DRS model, $ln[\psi|\theta^{S},\theta_{-S},U^{obs}]$ is not at all linear in $\theta^{S}$ and $\theta_{-S}$, where $\theta^{S}=N-x_{0}$, $\theta_{-S}=(\phi, p_{1.})$ and $\psi=p$.
%\paragraph{}
%The above two empirical Bayesian approaches EWiG and SEMWiG are presented here as an alternative solution to DA in the context of population size estimation under the model $M_{tb}$ when only two trapping samples are available. In next section, we consider various artificial populations reflecting different possible situations under the considered model and numerical results are given for the three said approaches under each of those populations.\par

\section{Numerical Illustrations}
\subsection{Simulation Study}
In this section we have studied the comparative performance of three different variants of Gibbs sampler for estimating human population size (\textit{N}) in DRS under the model $M_{tb}$, as described in last section. We consider four artificial populations characterized by different values of the DRS model parameters ($p_{1.}$, $p_{.1}$) and the same value of $N$ at 500; for one instances each of the two possible situations of behavioral effect - (\textit{i}) \textit{recapture averse} represented by $\phi=0.80$ and (\textit{ii}) \textit{recapture prone} represented by $\phi=1.25$. These four populations for each behavioral situation encompasses all possible combinations that occur in reality. All these eight combinations are presented in Table 2. The expected number of distinct captured individuals ($E(x_0)=N(p_{11}+p_{10}+p_{01})$) for different populations are also cited in that Table.
\begin{table}[ht]
\centering
\caption{Hypothetical populations with $N=500$ considered for simulations study}
\begin{minipage}{15cm}
\begin{tabular}{|ccccccccccc|}
\hline
Population & $\phi$ & $p_{1.}$ & $p_{.1}$&  $E(x_0)$ & & Population & $\phi$ & $p_{1.}$ & $p_{.1}$ & $E(x_0)$\\
\hline
\hline
P1 & 1.25 & 0.50 & 0.65 & 394  & &  P5 & 0.80 & 0.50 & 0.65 & 430\\
P2 & 1.25 & 0.60 & 0.70 & 422  & & P6 & 0.80 & 0.60 & 0.70 & 459\\
P3 & 1.25 & 0.80 & 0.70 & 458  & & P7 & 0.80 & 0.80 & 0.70 & 483\\
P4 & 1.25 & 0.70 & 0.55 & 420  & & P8 & 0.80 & 0.70 & 0.55 & 446\\
\hline
\end{tabular}
\end{minipage}
\label{tab:realresult}
\end{table}
It is noted that in the first two populations for each $\phi$, $p_{1.}<p_{.1}$ which refers to the usual situation in DRS data obtained by a specialised survey conducted after a large census operation, e.g. Post Enumeration Survey (PES). The last two populations with $p_{1.}>p_{.1}$ are just the opposite case which are also relevant in practice.
%We also include the Nour's (1982) estimator for DRS in our comparative study for $\phi=1.25$ case only as he deduced his approach for recapture prone situation.
\paragraph{}
200 data sets on ($x_{11},x_{01},x_{10}$) were generated from each population. Estimates from DA have been obtained through Gibbs sampling with adaptive rejection sampling (ARS) technique to draw samples for $p$ from (9), the conditional posterior density of $p$ (as according to our model, $\pi(p|N,\phi,\textbf{\underline{x}})$ does not have explicit form). We follow Lee et al. (2003, pp. 482) to implement ARS. Let $S_n$ = \{$p_1$, $p_2$,\ldots, $p_n$\}, where $p_1<p_2<\ldots<p_n$, denote a current set of abscissa in the range (0, 1). Considering $n=5$, the initial value of $p_3$ (middle component of the set of abscissa, $S_{n}$ for $n=5$) is chosen as ($\hat{c}/\phi$) from the relation $E(x_{11})=p\phi E(x_{1.})$. DA and SEMWiG are designed in a way so that convergence of their Gibbs sampler can be examine through $\hat{R}^{1/2}$. To compute $\hat{R}^{1/2}$ in multiple chain method for convergence diagnostic, we take 5 independent parallel chains and hence burn-in period \emph{k} is fixed at \emph{h} for which $\hat{R}_{h}^{1/2}$ becomes smaller than 1.1 and then almost stabilizes. Convergence of EWiG methods is judged by plotting $N^{(h)}$ against $h$. After fixing the burn-in $k$, next $k$ number of samples are used to construct the estimate of posterior densities. For all approaches, final estimate of \emph{N} is obtained by averaging 200 replicated estimates. Based on those 200 estimates, the sample s.e., sample RMSE (Root Mean Square Error) and coverage proportion (in $\%$) are also calculated. Additionally, we calculate the $95\%$ credible interval (C.I.) based on sample quantile of the marginal posterior distribution of \emph{N}. All the results for different variants of Gibbs samplers are summarized in Table $3-4$ (for $\phi=1.25$) and Table $5-6$ (for $\phi=0.80$). We use the estimates of parameters under independence assumption ($\phi=1$) as their initial values for EWiG-I, EWiG-II and SEMWiG. Based on availability of directional information, priors on $\phi$ are assigned to all methods except for EWiG-II. In EWiG-II method, under the assumption of $\phi>1$, $p$ is maximised over its relevant domain (0, $\hat{c}$) and for $\phi<1$, $p$ is maximised over ($\hat{c}$, 1).
%For Nour's approach, estimate as well as its S.E., RMSE, $95\%$ confidence interval are computed over 200 generated datasets and present them as average estimate, sample SE, sample RMSE, $95\% CI$ respectively in Tables 4 along with three other Bayes estimates.
\begin{table}[ht]
\centering
\caption{Summary results for all the discussed Gibbs sampling approaches applied to the populations P1-P4 when the information $\phi>1$ is available.}
\begin{minipage}{16.5cm}
\begin{tabular}{|lcrcccc|}
\hline
%&&& \multicolumn{4}{c|}{Population}\\
Meth. & Prior on $\phi$ & & P1 & P2 & P3 & P4\\
\hline
\hline
%\multicolumn{7}{|c|}{\underline{When the information $\phi>1$ is available}}\\
& & &  &  &  & \\
DA & U($1, 2$) & $\hat{N}$(s.e.) & 472(18.94) & 478(11.04) & 490(7.55) & 488(12.38)\\
 & & RMSE  & 34.11 & 24.33 & 12.02 & 17.03\\
  & & C.I.  & $(438, 518)$ & $(451, 519)$ & $(473, 515)$ & $(457, 531)$\\
   & & Coverage  & 76 & 92 & 92 & 97\\
%     & & k  & 200 & 200 & 200 & 200\\
\multicolumn{7}{|c|}{}\\
EWiG-I & U($1,p^{-1}$) & $\hat{N}$(s.e.) & 478(15.34) & 490(11.31) & 489(6.84) & 477(12.77)\\
 & & RMSE  & 26.41 & 15.30 & 12.51 & 26.53\\
  & & C.I.  & $(457, 503)$ & $(471, 510)$ & $(477, 505)$ & $(458, 500)$\\
   & & Coverage  & 60 & 82 & 66 & 50\\
%        & & Time(in sec.)  & 15 & 9 & 15 & 15\\
\multicolumn{7}{|c|}{}\\
EWiG-II & $\phi>1$ & $\hat{N}$(s.e.) & 467(15.92) & 472(12.19) & 487(7.69) & 485(14.25)\\
 & & RMSE  & 36.90 & 30.70 & 15.16 & 20.82\\
  & & C.I.  & $(447, 489)$ & $(456, 490)$ & $(475, 501)$ & $(464, 509)$\\
   & & Coverage  & 24 & 23 & 54 & 68\\
%        & & Time(in sec.)  & 4 & 4 & 4 & 3\\
\multicolumn{7}{|c|}{}\\
SEMWiG & U($1,p^{-1}$) & $\hat{N}$(s.e.) & 484(18.10) & 485(14.06) & 495(8.53) & 502(16.16)\\
 & & RMSE  & 23.92 & 21.25 & 9.87 & 16.32\\
  & & C.I.  & $(451, 522)$ & $(456, 511)$ & $(477, 512)$ & $(474, 532)$\\
   & & Coverage  & 98 & 86 & 88 & 100\\
%        & & Time(in sec.)  & 12 & 12 & 12 & 12\\
\hline
\end{tabular}
\end{minipage}
%\label{tab:realresult}
\end{table}
\begin{table}[ht]
\centering
\caption{Summary results for populations P1-P4 when \textbf{No} directional information on $\phi$ is available.}
\begin{minipage}{16.5cm}
\begin{tabular}{|lcrcccc|}
\hline
%&&& \multicolumn{4}{c|}{Population}\\
Meth. & Prior on $\phi$ & & P1 & P2 & P3 & P4\\
\hline
\hline
%\multicolumn{7}{|c|}{\underline{When no such information on $\phi$ is available}}\\
& & &  &  &  & \\
DA & U($0.5,2$) & $\hat{N}$(s.e.) & 468(20.56) & 483(18.45) & 485(6.61) & 471(8.11)\\
 & & RMSE  & 37.94 & 24.97 & 16.97 & 30.61\\
  & & C.I.  & $(398, 561)$ & $(426, 560)$ & $(460, 513)$ & $((422, 542)$\\
   & & Coverage  & 88 & 95 & 91 & 100\\
%     & & Time(in sec.)  & - & - & - & -\\
\multicolumn{7}{|c|}{}\\
EWiG-I & GB-I(1,1,$p$) & $\hat{N}$(s.e.) & 488(19.21) & 504(21.36)  & 507(10.41) & 533(20.80)\\
 & & RMSE & 33.61 & 21.90 & 12.81 & 39.80\\
  & & C.I. & $(460, 520)$ & $(480, 532)$ &  $(489, 529)$ & $(492, 568)$\\
   & & Coverage  & 72 & 79 & 92 & 65\\
%        & & Time(in sec.)  & k=3000 & 9 & 15 & 15\\
\multicolumn{7}{|c|}{}\\
EWiG-II & $\phi>0$ & $\hat{N}$(s.e.) & 456(16.17) & 463(12.03) & 480(7.13) & 476(13.40)\\
 & & RMSE  & 46.84 & 38.89 & 21.40 & 27.67\\
  & & C.I.  & $(439, 476)$ & $(450, 479)$ & $(470, 492)$ & $(457, 499)$\\
   & & Coverage  & 10 & 5 & 20 & 45\\
%        & & Time(in sec.)  & 15 & 9 & 15 & 15\\
\multicolumn{7}{|c|}{}\\
SEMWiG & GB-I(1,1,$p$) & $\hat{N}$(s.e.) & 459(13.25) & 487(13.45) & 496(7.72) & 458(9.65)\\
 & & RMSE  & 43.57 & 19.03 & 8.81 & 44.42\\
  & & C.I.  & $(431, 483)$ & $(459, 513)$ & $(481, 510)$ & $(438, 477)$\\
& & Coverage  & 60 & 89 & 98 & 30\\
%        & & Time(in sec.)  & 13 & 10 & 10 & 15\\
\hline
\end{tabular}
\end{minipage}
\label{tab:realresult}
\end{table}
\paragraph{}
When the knowledge $\phi>1$ is available (as in Table 3), DA performs better than EWiG-II in terms of RMSE and coverage proportion. It also produces relatively more efficient estimates than EWiG-I method when $p_{1.}>p_{.1}$. It is also noted that SEMWiG estimates are most efficient though their coverage are smaller than DA. DA with prior U(1,3) produces almost same result as U(1,2) but it has larger confidence length. Clearly, DA has the best coverage among all the different versions of Gibbs sampler. But in worst cases due to small capture probability (P1 and P4), length of DA is too wide. In good situations (P2 and P3), DA, EWiG-I and SEMWiG are closely comparable and among them SEMWiG is slightly better performed.
\begin{table}[ht]
\centering
\caption{Summary results for all the discussed Gibbs sampling approaches applied to the populations P5-P8 when the information $\phi<1$ is available.}
\begin{minipage}{16.5cm}
\begin{tabular}{|lcrcccc|}
\hline
%&&& \multicolumn{4}{c|}{Population}\\
Meth. & Prior on $\phi$ & & P5 & P6 & P7 & P8\\
\hline
\hline
%\multicolumn{7}{|c|}{\underline{When the information $\phi<1$ is available}}\\
& & &  &  &  & \\
DA & U($0.2, 1.5$)* & $\hat{N}$(s.e.) & 472(18.77) & 490(10.91) & 509(6.39) & 481(9.39)\\
 & & RMSE  & 35.42 & 13.65 & 10.34 & 20.08\\
  & & C.I.  & $(431, 565)$ & $(460, 560)$ & $(485, 545)$ & $(448, 545)$\\
   & & Coverage & 99 & 99 & 100 & 100\\
%     & & Time(in sec.)  & 200 & 200 & 200 & 200\\
\multicolumn{7}{|c|}{}\\
EWiG-I & U($\hat{c},1$) & $\hat{N}$(s.e.) & 530(19.03) & 529(12.45) & 517(7.07) & 520(12.39)\\
 & & RMSE  & 35.99 & 31.93 & 18.86 & 24.04\\
  & & C.I.  & $(506, 557)$ & $(510, 550)$ & $(504, 531)$ & $(499, 544)$\\
   & & Coverage  & 37 & 18 & 22 & 53\\
%        & & Time(in sec.)  & 2 & 2 & 2 & 2\\
\multicolumn{7}{|c|}{}\\
EWiG-II & $\phi<1$ & $\hat{N}$(s.e.) & 526(27.20) & 540(13.87) & 524(8.50) & 522(14.17)\\
 & & RMSE  & 37.08 & 45.50 & 27.10 & 26.17\\
  & & C.I.  & $(495, 558)$ & $(522, 565)$ & $(511, 542)$ & $(499, 548)$\\
   & & Coverage  & 60 & 7 & 9 & 55\\
%        & & Time(in sec.)  & 4 & 4 & 4 & 4\\
\multicolumn{7}{|c|}{}\\
SEMWiG & U($\hat{c},1$) & $\hat{N}$(s.e.) & 478(13.04) & 492(8.88) & 498(4.87) & 484(10.69)\\
 & & RMSE  & 25.27 & 11.74 & 5.28 & 19.34\\
  & & C.I.  & $(455, 507)$ & $(477, 510)$ & $(489, 508)$ & $(465, 506)$\\
   & & Coverage  & 60 & 86 & 73 & 69\\
%        & & Time(in sec.)  & 3 & 3 & 4 & 4\\
\hline
\end{tabular}
\end{minipage}
%\label{tab:realresult}
\end{table}
\begin{table}[ht]
\centering
\caption{Summary results for populations P5-P8 when \textbf{No} directional information on $\phi$ is available.}
\begin{minipage}{16.5cm}
\begin{tabular}{|lcrcccc|}
\hline
%&&& \multicolumn{4}{c|}{Population}\\
Meth. & Prior on $\phi$ & & P5 & P6 & P7 & P8\\
\hline
\hline
%\multicolumn{7}{|c|}{\underline{When no such information on $\phi$ is available}}\\
& & &  &  &  & \\
DA & U($0.5,2$) & $\hat{N}$(s.e.) & 474(20.80) & 512(15.76) & 516(6.17) & 517(13.02)\\
 & & RMSE  & 35.58 & 19.83 & 18.71 & 21.75\\
  & & C.I.  & $(431, 566)$ & $(461, 575)$ & $(486,553)$ & $(451, 615)$\\
   & & Coverage  & 99 & 100 & 100 & 100\\
%     & & Time(in sec.)  & - & - & - & -\\
\multicolumn{7}{|c|}{}\\
EWiG-I & GB-I(1,1,$p$) & $\hat{N}$(s.e.) & 495(23.46) & 532(29.65) & 540(18.95) & 569(29.20)\\
 & & RMSE  & 23.70 & 40.45 & 43.55 & 66.03\\
  & & C.I.  & $(463, 530)$ & $(500, 569)$ & $(512, 571)$ & $(519, 615)$\\
   & & Coverage  & 97 & 13 & 10 & 4\\
%        & & Time(in sec.)  & 4 & 4 & 4 & 4\\
\multicolumn{7}{|c|}{}\\
EWiG-II & $\phi>0$ & $\hat{N}$(s.e.) & 465(18.41) & 516(18.48) & 528(6.60) & 463(12.92)\\
 & & RMSE  & 39.87 & 24.25 & 29.42 & 39.62\\
  & & C.I.  & $(449, 486)$ & $(497, 538)$ & $(514, 546)$ & $(453, 477)$\\
   & & Coverage  & 22 & 50 & 2 & 10\\
%        & & Time(in sec.)  & 4 & 4 & 4 & 4\\
\multicolumn{7}{|c|}{}\\
SEMWiG & GB-I(1,1,$p$) & $\hat{N}$(s.e.) & 474(9.70) & 505(7.53) & 510(6.29) & 487(13.21)\\
 & & RMSE  & 27.75 & 8.91& 12.60 & 18.30\\
  & & C.I.  & $(454, 497)$ & $(487, 518)$ & $(498, 526)$ & $(466, 517)$\\
   & & Coverage  & 52 & 100 & 58 & 90\\
%        & & Time(in sec.)  & 12 & 13 & 13 & 15\\
\hline
\end{tabular}
\end{minipage}
%\label{tab:realresult}
\end{table}
\paragraph{}
When interest populations are P1 to P4 but the knowledge $\phi>1$ is not available (as in Table 4), DA performs better than EWiG-II in terms of RMSE and coverage proportion. It also produces relatively more efficient estimates than EWiG-I method only for P4. It is also noted that SEMWiG estimates are most efficient though their coverages are smaller than DA. Here also, DA has the best coverage among all the Gibbs sampling techniques. In this situations, EWiG-I shows best efficiency over all populations. In particular for healthy situation (as like P2 or P3), SEMWiG is also performed with similar kind of goodness as EWiG-I.
\paragraph{}
In Table 5, when the knowledge $\phi<1$ is available, both the methods in EWiG do not perform satisfactorily, they rather unanimously overestimate the population size for all populations (P5-P8).  Moreover, their confidence bounds also do not include the true population size 500 which results in small coverage. It is observed from Table 5 that here also SEMWiG works better than DA in terms of s.e. and RMSE but as coverage proportion is concerned, DA has more coverage of the true \textit{N} due to its wide confidence bound. One point is to be noted here that when we try to use prior U(0.2, 1) on $\phi$ as we know underlying $\phi$ is less than 1, then DA methods often take very very large time to generate samples due to its high density in conditional posterior even beyond the upper restriction 1. Hence, we use the prior U(0.2, 1.5) as another trial.
\paragraph{}
When data generated from the populations with $\phi<1$ (here, P5-P8) and the knowledge $\phi<1$ is not available, all methods generally overestimate the true \textit{N}. Table 6 shows us that EWiG-I performs very badly due to high overestimation. Other EWiG related method, EWiG-II, also produce significant overestimates. In this context DA and SEMWiG works quite similarly, producing slight overestimates for P6 and P7. DA is preferable in terms of coverage whereas SEMWiG may be the choice for its better efficiency.
\paragraph{}
\textbf{\textit{Remark.}} $\hat{R}_{h}^{1/2}$ stabilizes very fast ($h\geq150$) for all populations in case of DA with Lee et al.'s prior specification and we have shown the DA estimates against that \textit{h} for which $\hat{R}_{h}^{1/2}$ just cross 1.1. But Bayesian estimates, calculated for $h\geq150$, do not stabilize in parallel (\textit{see} Figure 1). However, the Bayesian estimates from DA (with Lee et al.'s prior specification) are ultimately converging at far above the true value ($N=500$) for $h\geq 400$. Reason behind that the original model has identifiability problem and fully Bayesian analysis is performed here using non-informative/flat priors. Hence, variability in the conditional posteriors may be high and as a result posteriors estimates become large. Whereas convergence of SEMWiG estimates are better and Figure 2 shows a good parity of them with the stabilization of $\hat{R}^{1/2}$ values.
\begin{figure}
  \centering
  \includegraphics[width=5in,height=7in]{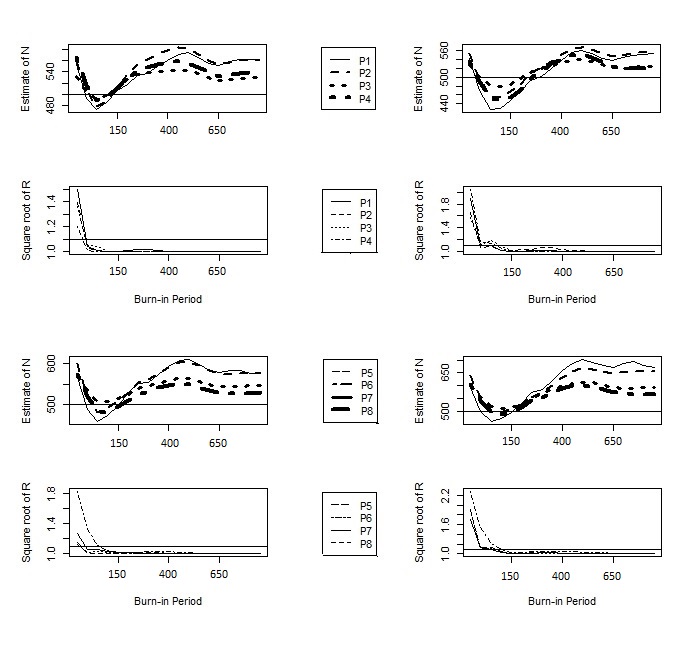}
  \caption{Plot of $\hat{N}$ (in 1st and 3rd row) and $\hat{R}^{1/2}$ (in 2nd and 4th row) against the index \emph{h} for \textbf{DA} method. True value of \textit{N} and suggested threshold value for $\hat{R}^{1/2}$ are indicated at 500 and 1.1 respectively. First two rows correspond populations \textbf{P1-P4} and last two rows for \textbf{P5-P8}. Plots in Left panel refers the situations of available directional knowledge on $\phi$ and right panel plots represent the situations with unavailable knowledge.}
\end{figure}
\begin{figure}
  \centering
  \includegraphics[width=5in,height=7in]{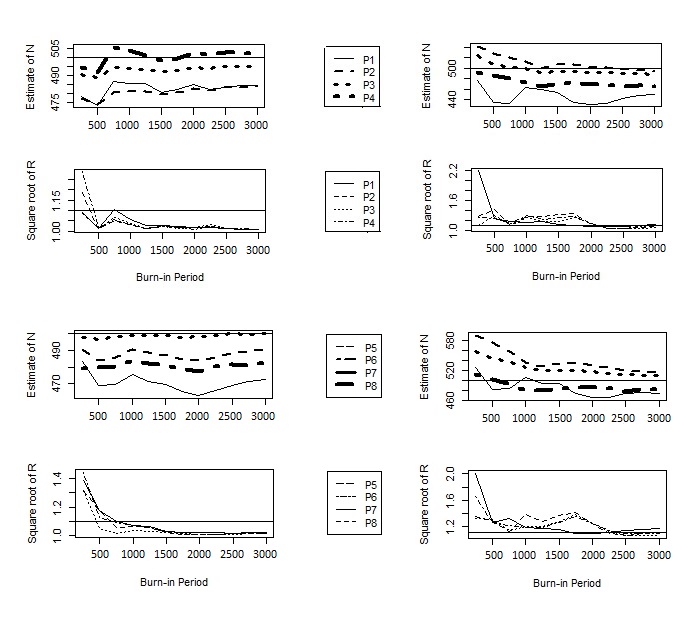}
  \caption{Plot of $\hat{N}$ (in 1st and 3rd row) and $\hat{R}^{1/2}$ (in 2nd and 4th row) against the index \emph{h} for \textbf{SEMWiG} method. True value of \textit{N} and suggested threshold value for $\hat{R}^{1/2}$ are indicated at 500 and 1.1 respectively. First two rows correspond populations \textbf{P1-P4} and last two rows for \textbf{P5-P8}. Plots in Left panel refers the situations of available directional knowledge on $\phi$ and right panel plots represent the situations with unavailable knowledge.}
\end{figure}

\subsection{Real Data Application}
Greenfield (1975) reports a DRS data on birth, death and migration obtained from a Population Change Survey conducted by the National Statistical Office in Malawi between 1970 and 1972. The sample was stratified into five areas: Blantyre, Lilongwe and Zomba urban areas; other urban areas and rural areas. However, in this article we choose the data on death records only for two strata - Lilongwe ($\hat{c}=0.593$, $x_{10}>x_{01}$) and Other urban areas ($\hat{c}=0.839$, $x_{10}<x_{01}$) due to its different \emph{c} values and opposite characters for $x_{10}$ and $x_{01}$. Nour (1982) estimated death sizes as \textit{378} and \textit{3046} for \textit{Lilongwe} and \textit{Other urban areas} respectively assuming the fact that two data sources are positively correlated in a human demographic study (i.e. in our context, $\phi>1$). Here, we employ the different versions of Gibbs sampling (discussed in section 3) and present the estimates of the death sizes for these two real populations under both the consideration of unrestricted $\phi$ (\textit{i.e.} $\phi>0$) and $\phi>1$. 200 parallel chains for $N$ are generated from different randomly selected starting points for $p_{1.}$, $\phi$ in DA. We therefore fix the burn-in period \emph{k} at 100 for \emph{Lilongwe} and 2500 for \emph{Other urban areas} after observing the plot of $\hat{R}_{h}^{1/2}$ with threshold value 1.1 in Figure 3 and record the remaining \emph{k} values. Similarly for SEMWiG method, burn-in period is fixed at 1100 for \emph{Lilongwe} and 9000 for \emph{Other urban areas} (\textit{see} Figure 3). This way the recorded values of (\emph{N}, \emph{$\phi$}) mimic their marginal posterior distributions. Summary of the analysis of this two populations by our suggested comparable methods are presented in Table 7.
\paragraph{}
Results against the four suggested priors for DA are given, of which first two are used when we don't have any information about the directional knowledge on $\phi$ and other two are presented considering $\phi>1$. For each of the other three methods EWiG-I, EWiG-II, and SEMWiG, first prior corresponds to the situation of no information about the directional knowledge on $\phi$ is available and the second one is for $\phi>1$. When no information on $\phi$ is available, DA says that the estimated number of deaths in \emph{Lilongwe} is 360 (with $95\%$ credible interval (348, 390)) and in \emph{Other urban areas} is around 3280 (with $95\%$ credible interval (2865, 3700)). DA with prior $U(0.5, 3)$ gives larger estimate for \emph{Other urban areas} but confidence interval is too wide. When recapture proneness is assumed, DA estimate the number of deaths at around 373 (with $95\%$ credible interval (357, 397)) and 3285 (with $95\%$ credible interval (2900, 4184)) for \emph{Lilongwe} and \emph{Other urban areas} respectively. For large population, DA estimate is lesser than Nour's. EWiG-I estimates become similar for both the different priors GB-I(1,1,$p$)$\equiv U(0, 1/p)$ and $U(1, 1/p)$ when population size is large. For small population (as \emph{Lilongwe}), estimate corresponding to $U(1, 1/p)$ has smaller C.V.. In summary, EWiG-I says that 370 and 3200 deaths occurred in \emph{Lilongwe} and \emph{Other urban areas} respectively, considering that these populations are recapture prone. When we also consider behavioral effect $\phi$ as another nuisance parameter, then the method EWiG-II estimates the true death counts for \emph{Lilongwe} and \emph{Other urban areas} at around 365 and 3200. For \emph{Lilongwe}, EWiG-II provides a lower estimate than any other method. For \emph{Other urban areas}, as \emph{N} is large, EWiG-I and EWiG-II become similar. Estimate from SEMWiG is slightly smaller than that of Nour (1982) and it gives around 372 and 2980 for the two said populations respectively. If the recapture proneness is not assumed, SEMWiG produces the estimates as 356 and 2845. One can observe that among all the competing methods, SEMWiG estimates have the lowest coefficient of variation (c.v.) and hence smaller length of confidence interval. With respect to the distributional variability, we can order the estimators as SEMWiG$<$EWiG-I$<$EWiG-II$<$DA. Note that DA, EWiG-II and SEMWiG suggest that \emph{Lilongwe} is not recapture prone.
\begin{figure}
  \centering
  \includegraphics[width=6in,height=7in]{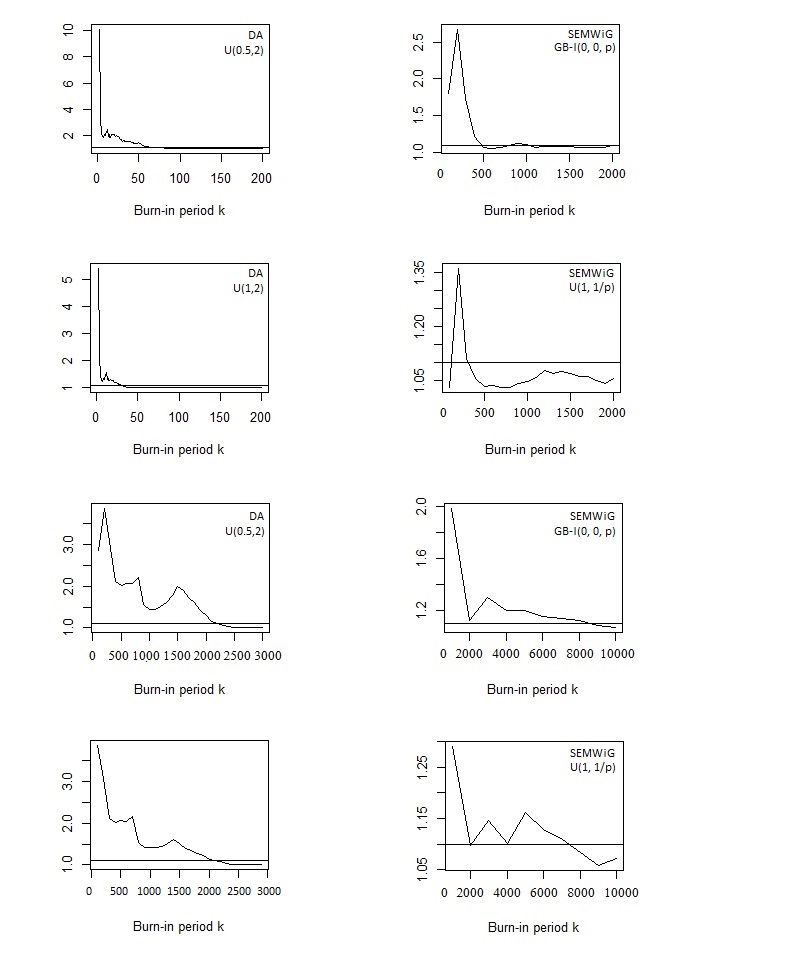}
  \caption{Plot of $\hat{R}^{1/2}$ against the index \emph{h}. Suggested threshold value of $\hat{R}^{1/2}$ is indicated at 1.1. First two rows are for \emph{Lilongwe} data and last two rows correspond to \emph{Other urban areas}. Methods with prior specifications are mentioned in right-top of each plot.}
\end{figure}
\begin{table}
\centering
\caption{Bayes and Empirical Bayes estimates of total number of deaths for different uniform priors on $\phi$. s.e. is computed based on sample posterior distribution and the $95\%$ posterior credible intervals for \textit{N} and \textit{$\phi$} is determined based on percentile method}
\begin{minipage}{15.5cm}
\begin{tabular}{|lcccccc|}
\hline
& & &  &   &  & \\
& $\pi(\phi)$ &  &  & $95\%$ CI  & & $95\%$ CI \\
Method & $/\Phi$ & $\hat{N}$(s.e.) & C.V.($\hat{N}$) & of \emph{N} & $\hat{\phi}$ & of $\phi$\\
\hline \hline
& & &  &   &  & \\
&\multicolumn{6}{c|}{Lilongwe} \\
& &  &  &  & & \\
DA & U(0.5, 2) & 359(12.51) & 0.035 & (348, 389) & 0.86 & (0.57, 1.62)\\
 &U(0.5, 3) & 362(12.84) & 0.035 & (348, 389) & 0.92 & (0.57, 1.62) \\
& U(1, 2) & 373(10.98) & 0.029 & (357, 397) & 1.22 & (1.01, 1.59)\\
& U(1, 3) & 373(9.65) & 0.026 & (357, 394) & 1.22 & (1.01, 1.64)\\
& &  &  &  & & \\
EWiG-I & GB-I(1,1,$p$) & 377(8.08) & 4.051 & (362, 394) & 1.29 & (1.11, 1.43)\\
 &U(1, $1/p$) & 370(6.35) & 3.185 & (359, 383) & 1.12 & (1.01, 1.22) \\
& &  &  &  & & \\
EWiG-II  & $\phi>0$ & 351(2.03) & 1.020 & (348, 356) & 0.67 & - \\
& $\phi>1$ & 365(5.22) &2.611 & (356, 376) & 1.01 & - \\
& &  &  &  & & \\
SEMWiG & GB-I(1,1,$p$) & 356(4.08) & 0.011 & (352, 362) & 0.79 & (0.68, 0.92)\\
 &U(1, $1/p$) & 372(1.30) & 0.004 & (370, 493) & 1.16 & (1.11, 1.20) \\
%& U(1, 2) & 374(4.76) & 0.013 & (371, 475) & 1.22 & (1.12, 1.49)\\
\hline
& &  &  &  &  & \\
&\multicolumn{6}{c|}{Other urban areas} \\
& &  &  &  & & \\
DA & U(0.5, 2) & 3276(218.75) & 0.067 & (2865, 3695) & 1.37 & (0.95, 1.79)\\
 &U(0.5, 3) & 3367(311.96) & 0.093 & (2898, 4184) & 1.46 & (0.98, 2.32) \\
& U(1, 2) & 3286(172.31) & 0.052 & (3005, 3647) & 1.38 & (1.09, 1.75)\\
& U(1, 3) & 3360(254.19) & 0.076 & (3008, 3986) & 1.46 & (1.09, 2.10)\\
& &  &  &  & & \\
EWiG-I & GB-I(1,1,$p$) & 3199(26.03) & 0.008 & (3149, 3251) & 1.29 & (1.26, 1.32)\\
 &U(1, $1/p$) & 3199(25.56) & 0.008 & (3151, 3247) & 1.29 & (1.27, 1.32)\\
& &  &  &  & & \\
EWiG-II & $\phi>0$ & 3198(25.50) &0.008 & (3150, 3249) & 1.29 & -\\
& $\phi>1$ & 3196(25.44) &0.008 & (3149, 3250) & 1.29 & - \\
& &  &  &  & & \\
SEMWiG & GB-I(1,1,$p$) & 2847(26.80) & 0.009 & (2818, 2885) & 0.93 & (0.89, 0.96)\\
 &U(1, $1/p$) & 2981(13.37) & 0.004 & (2970, 3001) & 1.06 & (1.05, 1.09) \\
%& U(1, 2) & 3059(45.79) & 0.015 & (3026, 3130) & 1.15 & (1.11, 1.22)\\
& &  &  &  &  & \\
\hline
\end{tabular}
\end{minipage}
\label{tab:realresult}
\end{table}
Moreover, it is also found that the people of \emph{Lilongwe} are less keen to give the information on deaths again in survey time than the \emph{Other urban areas} people. Over all, our suggested Bayes estimates with the assumption $\phi > 1$ are that around 372 deaths occurred in \emph{Lilongwe} and around 3200 deaths occurred in \emph{Other urban areas}.

\section{Summary and Conclusions}
Motivated by the extensive use of Dual-record system (DRS) in various real life practices in society, we have considered the problem of homogeneous population size estimation in a DRS framework where behaviour response effect plays a significant role along with time variation effect. We have viewed the problem as a missing data situation. The present model $M_{tb}$ suffers from identifiability problem where suitable Bayesian methods have potential to overcome that burden. As per knowledge, it is the first attempt to make inference in this complex DRS situation via Bayesian methodologies. We have presented two empirical Bayesian approaches along with a natural fully Bayesian method. One major issue here is whether incorporation of the prior knowledge on directional nature of $\phi$, if available, helps in producing better estimates. Here we sketch the different methodologies conditionally and unconditionally on the directional knowledge available on $\phi$. We have restricted ourselves to the use of non-informative priors so that subjectiveness can be reduced in the prior selection which leads to robustness of the inference.
\paragraph{}
In general it is seen that SEMWiG is overall the most efficient method (both in terms of RMSE and s.e.). On the other hand, DA provides better coverage than any other method but also it possesses lower efficiency in most situations than SEMWiG. In addition to that, the proposed DA with trial-and-error approach may take a long time to discover a suitable range for uniform prior for $\phi$. Sometimes it also fails to draw samples from its highly dispersed conditional posterior densities when true $\phi$ is closer to 1, when the knowledge of population proneness or averseness is available. We also observe that fully Bayesian analysis (DA) with non-informative and/or flat prior ultimately converges around an highly overestimated value in long iterations. Coming to the EWiG methods, EWiG-I performs better than EWiG-II in terms of efficiency and coverage in all situations except the populations P6-P8 with no prior information on $\phi$. Beside the computational advantage, SEMWiG is more transparent and relatively easy than EWiG to explain to the practitioner. Thus Bayesian analysis along with likelihood estimates (for nuisance parameters) helps to maintain small variability and converges around a reasonable value. Convergence of fully Bayesian analysis in this context may be better when suitable subjective priors are elicited. Thus, our proposed empirical Bayes methods are useful to obtain a more efficient estimate of population size from this complex DRS depending upon the various possible practical situations. Though the suggested empirical Bayes methods consume little more computational time than the fully Bayes method, they help to get rid of the present identifiability problem better and produce more efficient alternatives in most of the situations.
%%%%%%%%%%%%%%%%%%%%%%%%%%%%%%%%%%%%%%%%%%%%%%%%%%%%%%%%%%%%%%%%%%%%%%%%%%%%%%%%%%%%%%%%%%%%%%%%%%%%%%%%%%%%%%%%%%%%%%%%%%%%

\section*{Acknowledgment}
%This work is partially supported by the research fellowship award (No. $09/093(0125)/2010-EMR-I$) given to the first author from Council of Scientific and Industrial Research (CSIR), India.
We are grateful to Dr. Sourabh Bhattacharya for his valuable comments and suggestions which has helped us to improve the paper considerably.
%%%%%%%%%%%%%%%%%%%%%%%%%%%%%%%%%%%%%%%%%%%%%%%%%%%%%%%%%%%%%%%%%%%%%%%%%%%%%%%%%%%%%%%%%%%%%%%%%%%%%%%%%%%%%%%%%%%%%%%%%%%%

\section*{Appendix}
\paragraph{}
\textbf{A. \textit{Convergence of the posterior distribution in EWiG}}
\paragraph{}
In the general Bayesian setup (section 3), our model, treated as a missing data model, presents the hierarchy as \\
\begin{eqnarray}
U^{obs} &\sim & f(x|\theta,\psi), \theta=(\theta^S, \theta_{-S})\nonumber\\
\theta^S &\sim & \pi(\theta^S|\psi,\gamma)\\
\theta_{-S} &\sim & \pi(\theta_{-S}|\psi)\nonumber
\end{eqnarray}
In EWiG, the estimate of posterior density of $\theta_{-S}$ and $\theta^S$ are given in (12) and (13) respectively. Here, we are interested in the limiting behavior of these estimates as $M$ and $N \rightarrow \infty$. Let us define
\begin{eqnarray}
g(\psi;\psi') & = & \int_{\Theta^{S}}\pi(\theta_{-S}|\theta^{S},\psi,U^{obs})\pi(\theta^{S}|\psi',U^{obs})d\theta^{S}\nonumber\\
\hat{g}(\psi;\psi') & = & M^{-1}\sum_{j=1}^{M}\pi(\theta_{-S}|\theta^{S(j)},\psi,U^{obs})\nonumber
\end{eqnarray}
, where $\theta^{S(j)}\sim \pi(\theta^{S}|\theta_{-S}^{(j)},\psi',U^{obs})$. Since, our present study belongs to finite population statistics, hence the estimator $\hat{\psi}_{N}$ of nuisance parameter $\psi$ depends on \textit{N}. Now, the following lemma presents the conditions on which almost everywhere (a.e.) convergence of $\hat{\pi}(\theta_{-S}|\hat{\psi},U^{obs})$ to $\pi(\theta_{-S}|\psi,U^{obs})$ can be built.

\paragraph{}
\textbf{Lemma 1.} Let us consider $\psi_0$ as a true value of $\psi$. Under the following assumptions along with ergodicity of EM generated Markov chain:\\
A1. $\hat{\psi}_{N} \rightarrow \psi_0$ almost everywhere,\\
A2. $g(\psi;\psi')$ is continuous function of $\psi$ and $\psi'$,\\
A3. $\hat{g}(\psi;\psi')$ is continuous in $\psi$ and stochastically equicontinuous in $\psi'$,\\ \\
$\exists$ a sequence $M_N$ $\ni M_N \rightarrow \infty$ as $N \rightarrow \infty$, for which
\begin{equation}
\mid \hat{g}(\hat{\psi}_{N};\hat{\psi}_{N})-g(\psi_0;\psi_0)\mid  \stackrel{a.e.}{\rightarrow} 0 \mbox{as $N \rightarrow \infty$}
\end{equation}
\textit{Proof.} By triangle inequality we can write
\[ \mid \hat{g}(\hat{\psi}_{N};\hat{\psi}_{N})-g(\psi_0;\psi_0)\mid \leq \mid \hat{g}(\hat{\psi}_{N};\hat{\psi}_{N})-\hat{g}(\psi_0;\psi_0)\mid + \mid \hat{g}(\psi_0;\psi_0)-g(\psi_0;\psi_0)\mid.  \]
By Ergodic property, second term on right hand side converges to 0. For any given \textit{N}, choose $M_{N}^{(1)}$ so that $\mid \hat{g}(\psi_0;\psi_0)-g(\psi_0;\psi_0)\mid\leq\epsilon/3$. For first term, again by triangle inequality,
\[ \mid \hat{g}(\hat{\psi}_{N};\hat{\psi}_{N})-\hat{g}(\psi_0;\psi_0)\mid \leq \mid \hat{g}(\hat{\psi}_{N};\hat{\psi}_{N})-\hat{g}(\psi_0;\hat{\psi}_{N})\mid + \mid \hat{g}(\psi_0;\hat{\psi}_{N})-\hat{g}(\psi_0;\psi_0)\mid. \]
From assumption A3, the first term on right hand side in the above inequality tends to 0 as $N \rightarrow \infty$. Since $\hat{\psi}_{N} \rightarrow \psi_0$ a.e. by assumption A1, we can choose \textit{N} large enough so that $\mid\hat{\psi}_{N} \rightarrow \psi_0\mid<\delta$, except on a set with probability less than $\epsilon/2$. Assumption A3 also says that $\hat{g}(\psi;\psi')$ is stochastically equicontinuous in $\psi'$ which means, for given $\epsilon>0$, one can find a $\delta$($>$0) $\ni$ $\mid \hat{g}(\psi;\psi_1')- \hat{g}(\psi;\psi_2')\mid<\epsilon$ $\forall \mid \psi_1'-\psi_2'\mid<\delta$ except on a set with g-measure 0. Hence, we can choose $M_{N}^{(2)}$ so that $\mid\hat{g}(\psi_0;\psi')-\hat{g}(\psi_0;\psi_0)\mid\leq\epsilon/3$ $\forall \mid\psi'-\psi_0\mid<\delta$, except on a set with g-measure less than $\epsilon/2$. Hence, the second term in the right side is bounded as follows
\[ \mid \hat{g}(\psi_0;\hat{\psi}_{N})-\hat{g}(\psi_0;\psi_0)\mid \leq \stackrel[\psi':\mid \psi'-\psi_0\mid<\delta]{}{sup} \mid \hat{g}(\psi_0;\psi')-\hat{g}(\psi_0;\psi_0)\mid \leq \epsilon/3. \]
Thus for any arbitrary $\epsilon>0$, we can choose a large \textit{N} and $M_N=max(M_{N}^{(1)},M_{N}^{(2)})$. Hence, $\mid \hat{g}(\hat{\psi}_{N};\hat{\psi}_{N})-g(\psi_0;\psi_0)\mid\leq\epsilon$, except a set with probability less than $\epsilon/2+\epsilon/2=\epsilon$. Hence the proof of (5.2).$\Box$

\paragraph{}
\textbf{Lemma 2} (\textit{Scheff\'{e}'s Lemma.}) If $f_n$ is a sequence of integrable functions on a measure space ($X,\Omega,\mu$) that converges almost everywhere to another integrable function \textit{f}, then,\\
$\int \mid f_n(y)-f(y)\mid d\mu \rightarrow 0$ if and only if $\int \mid f_n\mid d\mu \rightarrow \int \mid f\mid d\mu$ for $n\rightarrow 0$.

\paragraph{}
\textit{Proof of Theorem 1}. Both the \textit{Lemma 1} and \textit{Lemma 2} together imply the \textit{Theorem 1}(\textit{1}) as $f_n=\hat{g}_{N}$ and $f=g$ both are density and hence positive. \textit{Theorem 1}(\textit{2}) can be proved following the same way just by exchanging $\hat{\pi}(\theta_{-S}|\hat{\psi},U^{obs})$ and $\pi(\theta_{-S}|\psi,U^{obs})$ in their places and defining
\begin{eqnarray}
g(\psi;\psi') & = & \int_{\Theta_{-S}}\pi(\theta^{S}|\theta_{-S},\psi,U^{obs})\pi(\theta_{-S}|\psi',U^{obs})d\theta_{-S}\nonumber\\
\hat{g}(\psi;\psi') & = & M^{-1}\sum_{j=1}^{M}\pi(\theta^{S}|\theta_{-S}^{(j)},\psi,U^{obs})\nonumber
\end{eqnarray}
, where $\theta_{-S}^{(j)}\sim \pi(\theta_{-S}|\theta^{S(j)},\psi',U^{obs})$.$\Box$
\\
\\
\textbf{B. \textit{Proof of Theorem 2}}
\paragraph{}
Two density functions F and G are said to be \textit{right tail equivalent} if they have the same right endpoints $\omega$ ($\leq\infty$) and $\stackrel[x\uparrow \omega]{}{lim} \frac{1-F(x)}{1-G(x)}=c$, for some constant $0<c<\infty$.\\
Consider $\theta^{S}=N-x_{0}$, $\theta_{-S}=(p_{1.},\phi)$ and $\psi=p$, hence for large \textit{N} and $M\rightarrow \infty$, \textit{Lemma 1} produces
\begin{eqnarray}
  \hat{\pi}(N|\hat{p},\textbf{\underline{x}})=M^{-1}\sum_{j=1}^{M}\pi(N|p_{1.}^{(j)},\phi^{(j)},\hat{p},\textbf{\underline{x}}) \hspace{0.1in} \stackrel{a.e.}{\rightarrow} \pi(N|p_0,\textbf{\underline{x}}).
\end{eqnarray}
%%%%%%%%%%%%%%%%%%%%%%%%%%%%%%%%%%%%%%%%%%%%%%%%%%%%%%%%%%%%%%%%%%%%%%%%%%%%%%%%%%%%%%%%%%%%%%%%%%%%%%%%%%%%%%%%%%%%%%%%%%%%%

\paragraph{}
\section*{References}
\begin{itemize}
%\item Bishop, Y., Fienberg, S. and Holland, P. (1975). \textit{Discrete Multivariate Analysis, Theory and Practice}. MIT Press, Cambridge, Massachusetts.
\item Ayhan, H. O. (2010). Estimators of vital events in dual-record systems. \textit{Journal of Applied Statistics} \textbf{27}, 157-169.
\item Boonstra, P. S., Mukherjee, B. \& Taylor, J. M. G. (2013). Bayesian Shrinkage Methods for Partially observed Data with Many Predictors. \textit{The Annals of Applied Statistics} \textbf{7}, 2272-2292.
\item Brittain, S. \& $B\ddot{ö}hning$, D. (2009). Estimators in capture–recapture studies with two sources. \textit{Advances in Statistical Analysis} \textbf{93}, 23-47.
%\item Casella, G. (2001). Empirical Bayes Gibbs sampling. \textit{Biostatistics} \textbf{2}, 485–500.
\item Castledine, B. J. (1981). A Bayesian Analysis of Multiple Recapture. \textit{J. Amer. Statist. Assoc.} \textbf{81}, 338-346.
%\item{Casella86} Casella, G. (1986). Stabilizing binomial n estimators. \textit{J. Amer. Statist. Assoc.} \textbf{81}, 172-175.
\item Celeux, G. \& Diebolt, J. (1985). The SEM Algorithm: a probabilistic teacher algorithm derived from the EM algorithm for the mixture problem. \textit{Computational Statistics Quaterly} \textbf{2}, 73-82.
%\item Diebolt, J. and Celeux, G. (1991). Asymptotic properties of a Stochastic EM Algorithm for estimating mixing proportions. Technical Report No. 228. Detartment of Statistics, University of Washington, USA.
%\item Celeux, G., Chauveau, D. and Diebolt, J. (1995). On Stochastic Versions of the EM Algorithm. Technical Report. INSTITUT NATIONAL DE RECHERCHE EN INFORMATIQUE ET EN AUTOMATIQUE.
\item ChandraSekar, C. \& Deming, W.E. (1949). On a method of estimating birth and death rates and the extent of registration. \textit{J. Amer. Statist. Assoc.} \textbf{44}, 101-115.
%\item Chao, A., Tsay, P. K., Lin, S. H., Shau, W. Y. and Chao, D. Y. (1996). Population Size Estimation for Capture–Recapture Models With Applications to Epidemiological Data. \textit{Proceedings of the American Statistical Association, Biometrics Section}, 108-117.
\item Chao, A., Chu, W. \& Chiu, H.H. (2000). Capture-Recapture when Time and Behavioral Response Affect Capture Probabilities. \textit{Biometrics} \textbf{56}, 427-433.
\item Chao, A., Tsay, P. K., Lin, S. H., Shau, W. Y. \& Chao, D. Y. (2001). Tutorial in Biostatistics: The Application of Capture-Recapture Models to Epidemiological Data. \textit{Statistics in Medicine} \textbf{20}, 3123-3157.
%\item{Cowles96} Cowles, M. K. and Carlin, B. P. (1996). Markov Chain Monte Carlo Convergence Diagnostics: A Comparative Review. \textit{J. Amer. Statist. Assoc.} \textbf{91}, 883-904.
%\item{Darr58} Darroch, J. N. (1958). The multiple recapture census, I. Estimation of closed population. \textit{Biometrika} \textbf{48}, 343-359.
%\item{Elliot00} Elliot, M. R. and Little, R. J. A. (2000). A Bayesian Approach to Combining Information from a Census, a Coverage Measurement Survey, and Demographic Analysis. \textit{J. Amer. Statist. Assoc.} \textbf{95}, 351-362.
\item Elliot, M. R. \& Little, R. J. A. (2005). A Bayesian Approach to 2000 Census Evaluation Using ACE Survey Data and Demographic Analysis. \textit{J. Amer. Statist. Assoc.} \textbf{100}, 380-388.
\item Fienberg, S. E. (1972). The Analysis of Incomplete Multi-Way Contingency Tables. \textit{Biometrics}, \textbf{28}, 177-202.
\item Gelman, A. (1996). \textit{Inference and Monitoring Convergence}. In \textit{Markov Chain Monte Carlo in practice} (Edited by W. R. Gilks, S. Richardson and D. J. Spiegelhalter). Chapman and Hall, London.
%\item{George90} George, E. I. and Robert, C. P. (1990). Capture-recapture models and Bayesian sampling. Technical Report No. 435. Department of Statistics, Stanford University, Stanford, California.
%\item{Geman84} Geman, S. and Geman, D. (1984). Stochastic relaxation, Gibbs distributions, and the Bayesian restoration of images. \textit{IEEE Transactions on Pattern Analysis and Machine Intelligence PAMI} \textbf{6}, 721–741.
\item George, E. I. \& Robert, C. P. (1992). Capture-recapture estimation via Gibbs sampling. \textit{Biometrika} \textbf{79}, 677-683.
\item Greenfield, C. C. (1975). On the estimation of a missing cell in a $2 \times 2$ contingency table. \textit{J. R. Statist. Soc. A} \textbf{138}, 51-61.
\item Gilks, W. R., Richardson, S. \& Speigelhalter, D. J. (1996). Markov Chain Monte Carlo in practice, First edn, Chapman \& Hall, London.
%\item Hook, E. B. and Regal, R. R. (1995). Capture-recapture methods in epidemiology: Methods and limitations. \textit{Epidemiologic Review} \textbf{17}, 243-264.
%\item Huggins, R. (1989). On the statistical analysis of capture-recapture experiments. \textit{Biometrika} \textbf{76}, 133-140.
%\item{El-Khorazaty75} El-Khorazaty, M. N. (1975). \textit{Methodological strategies for the analysis of categorical data from multiple-record systems}. An unpublished Ph.D. thesis submitted to Department of Biostatistics, University of North Carolina at Chapel Hill. Institute of Statistics Mimeo Series No. 1019.
%\item Jabine T. B. and Bershad, M. A. (1968). Some Comments on the Chandrasekar and Deming Technique for the Measurement of Population Change. A paper proposed for the CENTO symposium on Demographic Statistics, Karachi, Pakistan.
%\item Laplace, P. S. (1786). Sur les Naissances, les Mariages et les Morts. \textit{Histoire de L'Academie Royale des Sciences}, 1783, Paris.
\item Lee, S. M. \& Chen, C. W. S. (1998). Bayesian inference of Population Size for behavioral response models. \textit{Statistica Sinica} \textbf{8}, 1233-1247.
\item Lee, S. M., Hwang, W. H. \& Huang, L. H. (2003). Bayes estimation of Population Size from Capture-recapture Models with Time Variation and Behavior response. \textit{Statistica Sinica} \textbf{13}, 477-494.
%\item{Liang95} Liang, K. Y. and Zeger, S. L. (1995). Inference based on estimating functions in the presence of nuisanceparameters (with discussion). \textit{Statistica Sinica} \textbf{10}, 158-199.
\item Lloyd, C. J. (1994). Efficiency of Martingle Methods in Recapture Studies. \textit{Biometrika} \textbf{81}, 305-315.
%\item{Marks74} Marks, E.S., Seltzer, W. and Krotki, K. R. (1974). \textit{Population Growth Estimation}. The Population Council, New York.
%\item{Link03} Link, W. A. (2003). Nonidentifiability of Population Size from Capture-Recapture Data with Heterogeneous Detection Probabilities. \textit{Biometrics}, \textbf{59}, 1123-1130.
\item Nour, E. S. (1982). On the Estimation of the Total Number of Vital Events with Data from Dual-record Collection Systems. \textit{J. R. Statist. Soc. A} \textbf{145}, 106-116.
\item Otis, D. L., Burnham, K. P., White, G. C. \& Anderson, D. R. (1978). Statistical Inference from Capture Data on Closed Animal Populations. \textit{Wildlife Monographs }\textbf{62}, 1-135.
%\item{Raftery88} Raftery, A. E. (1988). Inference for Binomial N parameter: A Hierarchical Bayes Approach. \textit{Biometrika} \textbf{75}, 223-228.
%\item{Robert67} Roberts, H. V. (1967). Informative stopping rules and inferences about population size. \textit{J. Amer. Statist. Assoc.} \textbf{62}, 763-775.
%\item{Seber82} Seber, G. A. F. (1982). \textit{The Estimation of Animal Abundance and Related Parameters}. 2nd edition. Charles W. Griffin, London.
\item Seber, G. A. F. (1986). A Review of Estimating Animal Abundance. \textit{Biometrics} \textbf{42}, 267-292.
%\item Seber, G. A. F. (1992). A Review of Estimating Animal Abundance II. \textit{International Statistical Review}, \textbf{60}, 129-166.
%\item{Smith88} Smith, P. J. (1988). Bayesian Analysis for multiple capture-recapture surveys. \textit{Biometrics} \textbf{44}, 1177-1189.
\item Smith, P. J. (1991). Bayesian Analysis for a multiple capture-recapture model. \textit{Biometrika} \textbf{78}, 399-407.
\item Tanner, M. A. \& Wong, W. H. (1987). The calculation of posterior distributions by data augmentation. \textit{J. Amer. Statist. Assoc.} \textbf{82}, 528–550.
%\item Van Dyk, D. A. and MENG, X. L. (2001). Discussion Article - The Art of Data Augmentation. \textit{Journal of Computational and Graphical Statistics} \textbf{10}, 1-50.
\item Wei, G. C. G. \& Tanner, M. A. (1990). A Monte Carlo implementation of the EM algorithm and the poor man’s data augmentation algorithms. \textit{J. Amer. Statist. Assoc.} \textbf{85}, 699–704.
\item Wolter, K. M. (1986). Some Coverage Error Models for Census Data. \textit{J. Amer. Statist. Assoc.} \textbf{81}, 338-346.
%\item{Wu83} Wu, C. (1983). On the convergence properties of the EM algorithm. \textit{Annals of Statistics} \textbf{11}, 95-103.
\item Yang, H. C. \& Chao, A. (2005). Modelling Animals' Behavioral Response by Markov Chain Models for Capture-Recapture Experiments. \textit{Biometrics} \textbf{61}, 1010-1017.
\end{itemize}
\end{document}